\documentclass[prl,aps]{revtex4}  
\usepackage{graphicx}
\usepackage{amsmath}
\usepackage{amssymb}  
                                                          
\begin{document}    
\DeclareGraphicsExtensions{.eps}  
\title{Prediction of a surface state and a related surface insulator-metal
  transition for the (100) surface of stochiometric EuO} 
\author{R.~Schiller and W.~Nolting} 
\affiliation{Humboldt-Universit\"at zu Berlin, Institut f\"ur Physik,
         Invalidenstra{\ss}e 110, D-10115 Berlin, Germany}
\date{\today} 

\begin{abstract}
We calculate the temperature and layer-dependent electronic structure of
a 20-layer EuO(100)-film using a combination of first-principles and
model calculation based on the ferromagnetic Kondo-lattice model. The
results suggest the existence of a EuO(100) surface state which can lead
to a surface insulator-metal transition.
\end{abstract}
\maketitle

In the recent past many theoretical and experimental research works have been
focussed on the intriguing properties of rare earth metals and their
compounds. Among others, the extraordinary surface magnetic properties
of the lanthanides \cite{DMIL97}, as e.\,g.\  an enhanced Curie
temperature of the Gd(0001) surface compared to that of bulk Gd \cite{WAG+85},
have provoked numerous research activities. 
Concerning the interplay between electronic structure and 
exceptional magnetic properties at the Gadolinium surface 
a Gd(0001) surface state \cite{WF91,LHD+91}
is believed to play a crucial role and its temperature dependent
behaviour has been discussed intensely, e.\,g,\  
\cite{DDN98}. 

Rare-earth materials are so-called local-moment systems, i.e.~the
magnetic moment 
stems from the partially filled {\em 4f}-shell of the rare-earth atom
being strictly localized at the ion site. Thus
the magnetic properties of these materials are determined by the
localized magnetic moments. On the other hand, the electronic properties like
electrical conductivity are borne by itinerant electrons in rather broad
conduction bands, e.g.~{\em 6s}, {\em 5d} for Gd. Many of the characteristics of
local-moment systems can be attributed to a correlation between the localized
moments and the itinerant conduction electrons. For this situation the
ferromagnetic Kondo-lattice model (FKLM) which is also referred to as
the {\em s-f} model has been proven to be an adequate description. In
this model, the correlation  between localized moments and conduction
electrons is represented by an intraatomic exchange interaction.

In this work we will introduce a multi-band FKLM and use it to
calculate the temperature and layer-dependent bandstructure of EuO
films. For all the calculations, the surfaces of the films are parallel
to the fcc(100) crystal plane and the films will be referred to as
EuO(100) films. These films consist of $n$ equivalent 
parallel layers. The lattice sites within the film are indicated by a greek
letter $\alpha$, $\beta$, $\gamma$, $\ldots$, denoting the layer, and by a latin
letter $i$, $j$, $k$, $\ldots$, numbering the sites within a given
layer. The different subbands of the conduction bands of EuO will be
denoted by the indices $m$, $m'$. 

The Hamiltonian for the multi-band FKLM, 
\begin{equation}
\label{ham}
{\cal H} = {\cal H}_s + {\cal H}_{f} + {\cal H}_{sf},
\end{equation}
consists of three parts. The first,
\begin{equation}
\label{h_s}
{\cal H}_s = \sum_{ij\alpha\beta\sigma} T^{mm'}_{ij\alpha\beta} 
c^+_{i\alpha m\sigma} c_{j\beta m'\sigma},
\end{equation}
contains the kinetic energy of the conduction band electrons.
$c^+_{i\alpha m\sigma}$ ($c_{i\alpha m\sigma}$) is the creation (annihilation)
operator of an electron with spin $\sigma$ from the $m$-th subband 
at the lattice site ${\bf R}_{i\alpha}$. 
The $T^{mm'}_{ij\alpha\beta}$ describe the hopping between the $m$-th
subband at the lattice site ${\bf R}_{i\alpha}$ and the $m'$-th subband
at the lattice site ${\bf R}_{j\beta}$. These hopping integrals have to
be determined within a first-principles band-structure calculation.

The second part of the Hamiltonian represents the system of the localized 
{\em f}-moments and consists itself of two parts,
\begin{equation}
\label{h_f}
{\cal H}_f = -\sum_{ij\alpha\beta} J^{\alpha\beta}_{ij} {\bf S}_{i\alpha}
{\bf S}_{j\beta} -D_0\sum_{i\alpha}\left(S^z_{i\alpha}\right)^2,
\end{equation}
where the first is the well-known Heisenberg interaction. Here the 
${\bf S}_{i\alpha}$ are the spin operators of the localized magnetic moments,
which are coupled by the exchange integrals, $J^{\alpha\beta}_{ij}$.
The second contribution is a single-ion anisotropy term which arises from the
necessity of having a collective magnetic order at finite temperatures, 
$T>0$ \cite{SN99a}. 
The according anisotropy constant $D_0$ is typically smaller by some
orders of magnitude than the Heisenberg exchange integrals, 
$D_0\ll J^{\alpha\beta}_{ij}$.

In addition to the contribution of the itinerant-electron system and the
contribution of the localized {\em f}-moments we have a third contribution to
account for an intraatomic interaction between the conduction electrons and the
localized {\em f}-spins. The form of that contribution in the case of
multiple conduction bands can be derived from the general form of the
on-site Coulomb interaction between electrons of different subbands,
with the restriction that electron scattering processes caused by the
Coulomb interaction are restricted to two involved subbands\cite{Schiller00}. 
By distinguishing between bands of localized and itinerant electrons,
respectively, one gets the on-site Coulomb interaction between both
groups of electrons. For the special situation of europium with an exactly
half-filled localized {\em 4f}-shell and by assuming the inter-band
exchange to be independent on the different subbands, one eventually
gets the multi-band {\em s-f} interaction:
\begin{equation}
  \label{eq:h_sf_multi}
  {\cal H}_{sf} = - \frac{J}{\hslash} \sum_{i\alpha m} 
  \boldsymbol{\sigma}_{i\alpha m} \cdot {\bf S}_{i\alpha}.
\end{equation}
where ${\bf S}_{i\alpha}$ is the local moment originating from the
Hund's rule coupling of the localized {\em f}-electrons,
$\boldsymbol{\sigma}_{i\alpha m}$ is the Pauli spin operator of an
electron from the $m$-th band of itinerant electrons, both at the
lattice site ${\bf R}_{i\alpha}$, and $J$ is the
intraatomic {\em s-f} exchange interaction. 

The many-body problem that arises with the Hamiltonian (\ref{ham}) is far from
being trivial and a full solution is lacking even for the case of the bulk.
In a previous paper we have presented an approximate treatment of the special
case of a single electron in an otherwise empty conduction band
\cite{SN99b}. The presented approach applies to
the situation in a model-type ferromagnetic semiconductor film
containing a single-nondegenerated {\em s}-like band. However, the
approach as well as the obtained results are directly transferable to
the case of the multi-band Kondo lattice model applicable to a real
ferromagnetic semiconductor film like ${\rm EuO}$.

\begin{figure}[t!]
  \centerline{\includegraphics[width=0.9\linewidth]{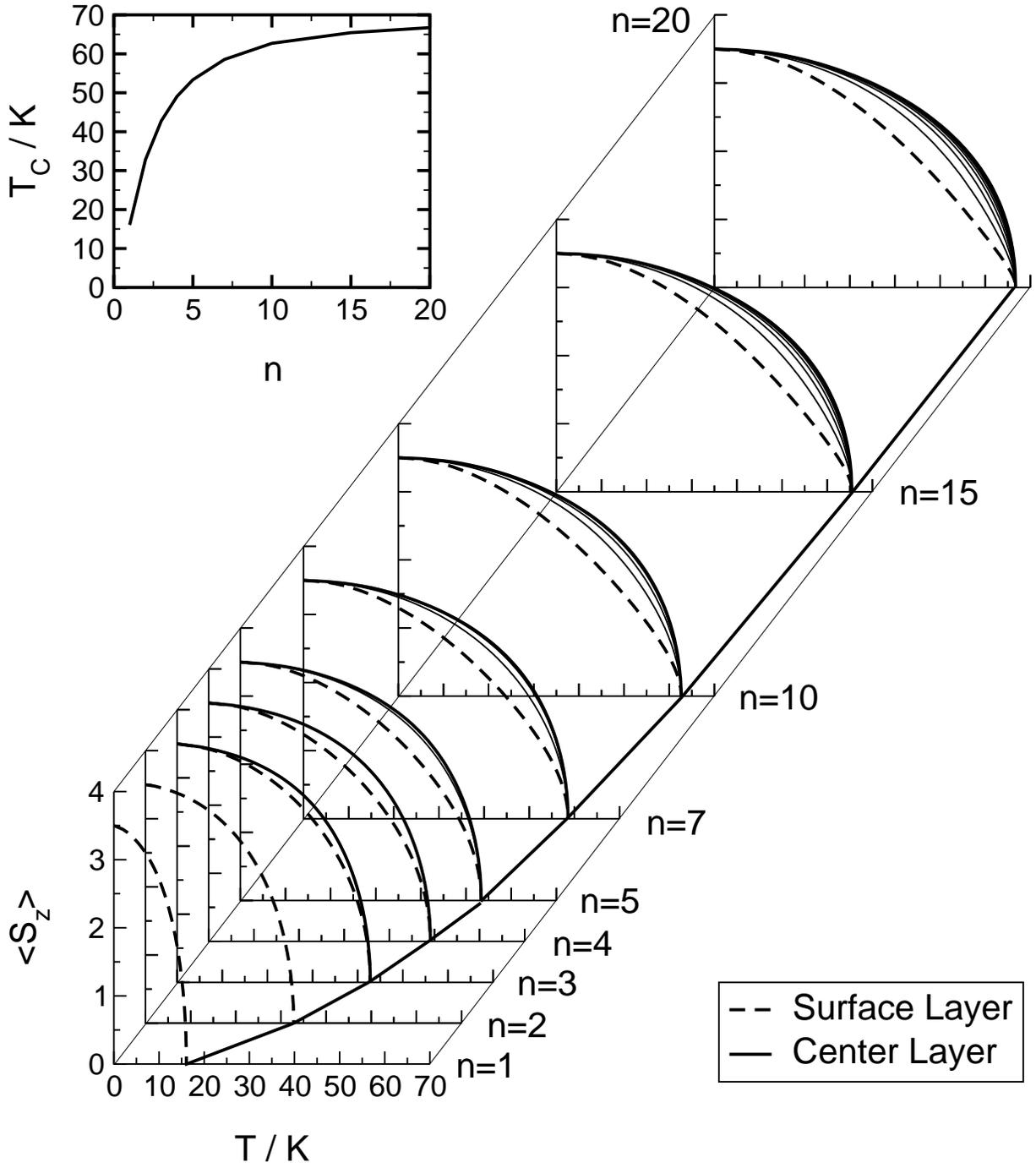}}
  \caption{Layer-dependent magnetizations, 
    $\langle S^z_{\alpha}\rangle$, of ${\rm EuO}(100)$ films as a
    function of temperature for $J_1/k_{\rm B}=0.625\,{\rm K}$, 
    $J_2/k_{\rm B}=0.125\,{\rm K}$, $D_0/k_{\rm B}=0.05\,{\rm K}$, for
    various thicknesses $n$ of the films. 
    {\bf Inset:} Curie temperatures as a function of film thickness.}
  \label{figure1}
\end{figure}

Due to the empty conduction bands we are considering throughout the
whole paper, the Hamiltonian  (\ref{ham}) can be split into an
electronic part, ${\cal H}_s + {\cal H}_{sf}$, and a magnetic part,
${\cal H}_f$, which can be solved separately \cite{SN99b}. 
For the magnetic subsystem ${\cal H}_f$ we employ an approach based on
the random phase approximation (RPA) which has been described in detail
in \cite{SN99a}. As the result one gets the layer- and temperature
dependent magnetizations of the local-moment system. These are depicted
for EuO(100) films of various thicknesses in Fig.~\ref{figure1}.
For the respective calculations, the Heisenberg interactions have been
approximated within the tight-binding approximation, taking into account
the nearest and the next nearest neighbour interactions, with values for
EuO of $J_1/k_B=0.625\,{\rm K}$ and $J_2/k_B=0.125\,{\rm K}$,
respectively \cite{quantmag2}. The calculated Curie temperature
converges as a function of film thickness in a manner similar to that
found experimentally for Gd(0001) films \cite{FBS+93}. Moreover, the
calculated Curie temperature of a 20-layer EuO(100) film of 
$66.7\,{\rm K}$ approximates very well the experimental value of 
$T_{\rm C}$ for bulk EuO of $69.33\,{\rm K}$ \cite{Wac91}.

For the electronic part a moment-conserving decoupling approximation
(MCDA) for suitably defined Green functions is employed \cite{SN99b}.
In the set of equations which constitute the solution for the electronic
subsystem, there appear {\em f}-spin correlation functions which have to
be evaluated within the magnetic subsystem (Eq.~(\ref{h_f}). 
These {\em f}-spin correlation functions mediate the temperature
dependence of the electronic structure of the system.

For the limiting case of ferromagnetic saturation of the localized 
{\em f}-spin system ($T=0$), there exists an {\em exact} solution for
the FKLM with empty conduction band ($n=0$)
\cite{SMN97}. While the spin-$\downarrow$ spectrum exhibits strong
correlation effects, the spectrum of the spin-$\uparrow$
electron is only rigidly shifted compared to the free solution obtained
for the case of vanishing {\em s-f} interaction, $J=0$. This means, that
the spin-$\uparrow$ spectra obtained within a ferromagnetic
band-structure calculation provide the single-particle input
(cf.~Eq.~(\ref{h_s})) for the FKLM model calculation.
Since the solution of the FKLM for $T=0$ and $n=0$ is exact, 
the problem of double counting of relevant interactions, which would
normally occur when superimposing the results of a band-structure calculation
with a model calculation, is elegantly avoided in the presented approach. 

To obtain the temperature-dependent band structure of a EuO(100) film,
the $T=0$ band structures are needed as single-particle input for the
model calculations. These have been calculated for the 
Eu-{\em 5d} bands, which in EuO dominate the conduction band, using a
standard TB-LMTO program \cite{And75}.
To account for the film geometry, we defined a supercell 
consisting of 20 EuO(100) layers followed by 5 layers of empty spheres, all
equidistantly spaced. The lattice constant 
for the EuO(100) film has been choosen to be the lattice constant of
bulk EuO, $a=5.142\,{\rm \AA}$ \cite{Wac91}. In this approach we neglect
surface relaxation and reconstruction, which might be present in EuO
films. 

\begin{figure}[t!]
  \centerline{\includegraphics[width=0.9\linewidth]{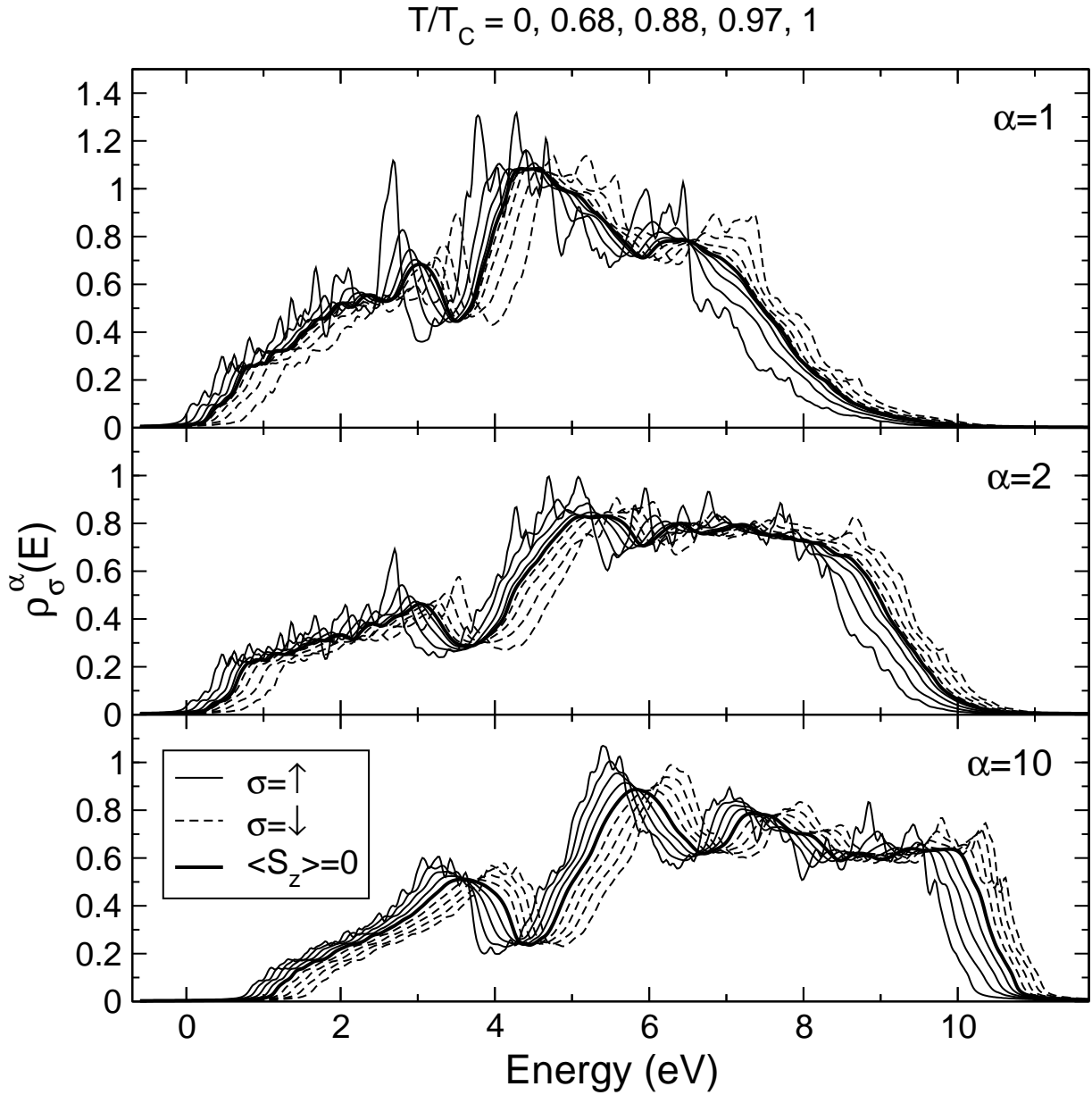}}
  \caption{Local densities of states of the empty 
    ${\rm Eu}$-{\em 5d}\/ bands of the first ($\alpha=1$), second
    ($\alpha=2$) and center ($\alpha=10$) layer of 
    a 20-layer ${\rm EuO}(100)$ film for different temperatures
    ($T_{\rm C}=66.7\,{\rm K}$). The energy zero refers to the Fermi
    energy \cite{Fermi_energy}.}  
  \label{figure2}
\end{figure}

\begin{figure}[t!]
  \centerline{\includegraphics[width=\linewidth]{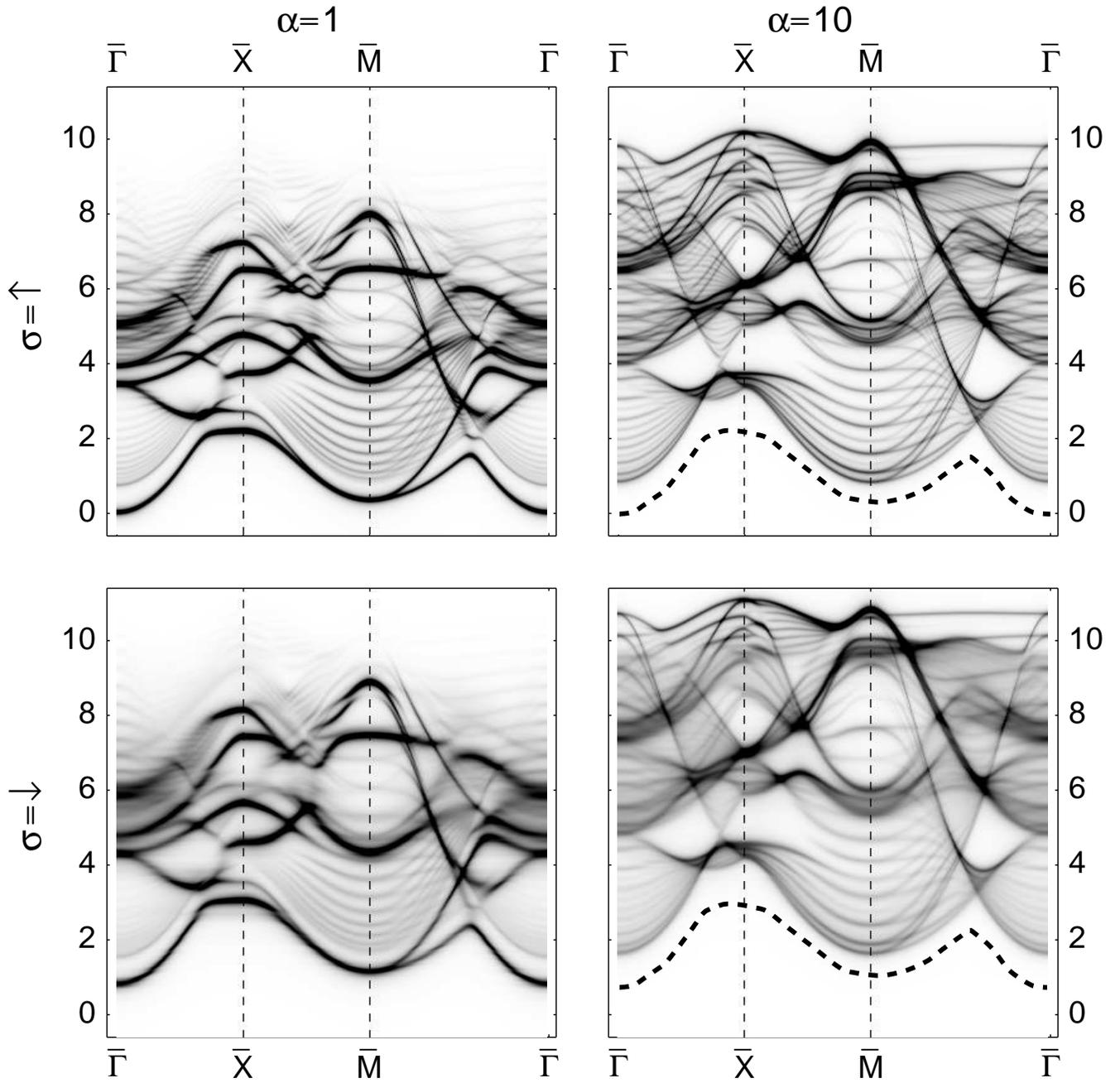}}
  \caption{Local spectral densities $S^{\alpha\alpha}_{{\bf k}\sigma}$ of
    the surface ($\alpha=1$) and the center layer ($\alpha=10$) of a
    20-layer ${\rm EuO}(100)$ film for $T=0$ and $J=0.25\,{\rm eV}$.
    The dashed lines in the spectral densities for the center layers
    (right hand side) reproduce the positions of the lower band edges of
    the surface state bands as given by the spectral densities of the
    surface layers (left hand side). The energy zero refers to the Fermi
    energy.}   
  \label{figure3}
\end{figure}

\begin{figure}[t!]
  \centerline{\includegraphics[width=\linewidth]{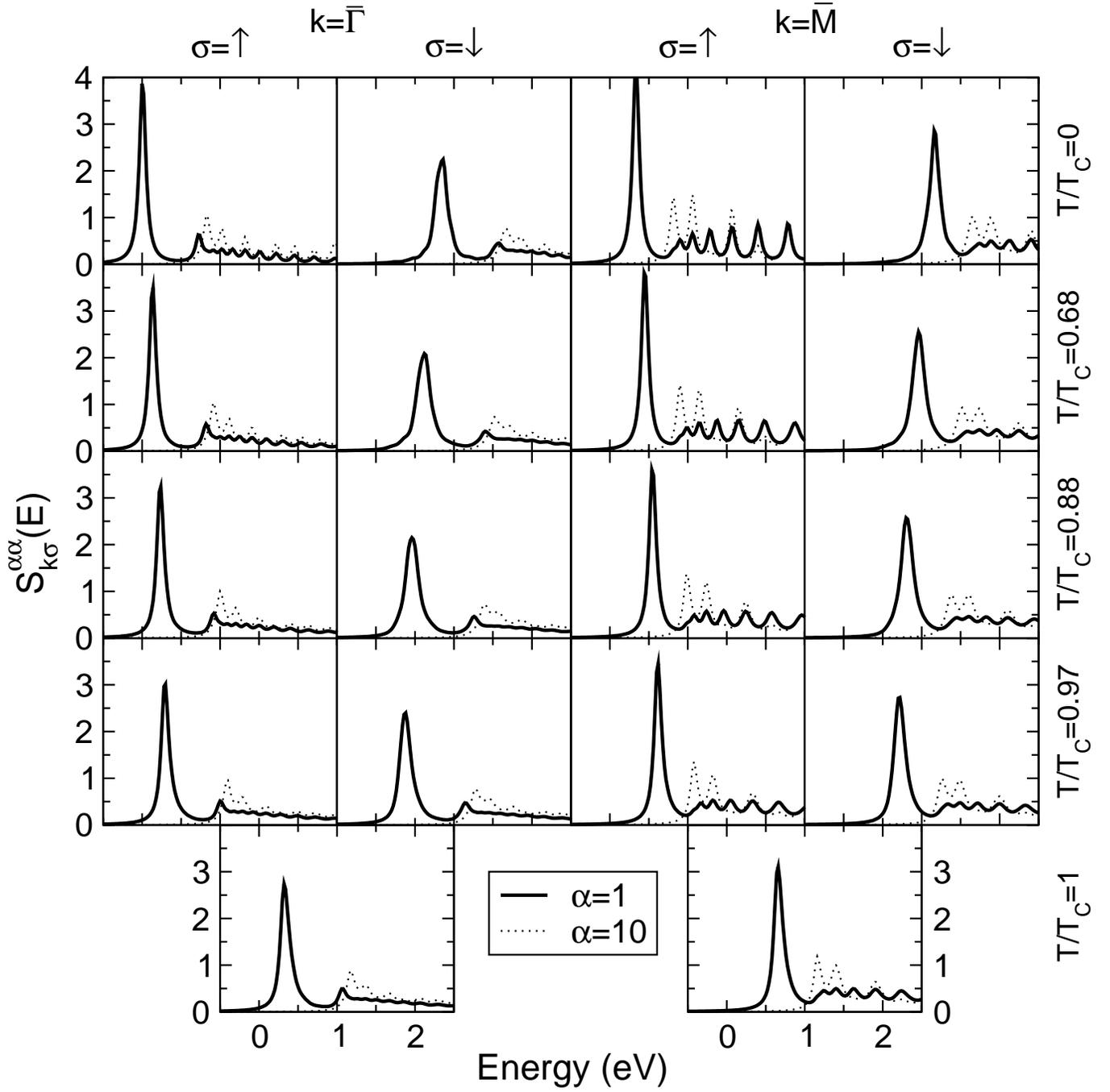}}  
  \caption{Local spectral densities $S^{\alpha\alpha}_{{\bf k}\sigma}$ of
    the surface ($\alpha=1$, solid lines) 
    and the center layer ($\alpha=10$, dotted lines) of a
    20-layer ${\rm EuO}(100)$ film at the $\Gamma$-point and the M-point
    for $J=0.25\,{\rm eV}$ and for different temperatures
    ($T_{\rm C}=66.7\,{\rm K}$). The energy zero refers to the Fermi energy.}  
  \label{figure4}
\end{figure}

Fig.~\ref{figure2} shows the temperature- and spin-dependent local
densities of states of the first, the second and the center layer of a
20-layer EuO(100) film with an {\em s-f} exchange interaction of 
$J=0.25\,{\rm eV}$. Here, as in Figs.~\ref{figure3} and \ref{figure4}
the energy zero refers to the Fermi energy of the system \cite{Fermi_energy}. 
For $T=0$, the spin-$\uparrow$ and the
spin-$\downarrow$ spectra occupy the lowest and the highest energy
range, respectively. With increasing temperature, the spectra for the
two spin-directions gradually approach each other until for $T=T_C$ 
($\langle S_z \rangle / S=0$, fat line) the spectra for both spin-directions
become the same. As a result, the lower band edge of the spin-$\uparrow$
bands is shifted towards lower energies when decreasing the
temperature below $T_C$. This effect is known as the red shift of the
optical absorption edge in bulk EuO \cite{Wac91}.
The spin-$\uparrow$ densities of state for $T=0$ 
($\langle S_z \rangle / S=1$) are, except for constant energy shift,
those obtained within the band-structure calculations, which is due to
the lack of possibility for the spin-$\uparrow$ electron to exchange its
spin with the system of perfectly aligned local moments. However, the
spin-$\downarrow$ electron can flip its spin even for $T=0$ by emitting
a magnon and forming a polaron-like quasiparticle \cite{SN99b}. Hence, the
spin-$\downarrow$ spectra for $T=0$ already exhibit correlation effects 
and differ from those of the band-structure calculations. 

In Fig.~\ref{figure2} the lower band edges
of the local densities of states of the surface layer ($\alpha=1$) 
lie at lower energies than the respective local densities of state 
of the center layers and are even touching the Fermi energy, already
indicating the existence of a surface state band at the lower edge of the 
${\rm Eu}$-{\em 5d}\/ bands and a possible surface insulator-metal
transition. In the following, these interesting indicative observations
shall be investigated in more detail.

In Fig.~\ref{figure3} the $T=0$ local spectral density of the
surface ($\alpha=1$) and the center ($\alpha=10$) layer of a 20-layer
${\rm EuO}(100)$ film can be seen for both spin directions. 
A surface state is a state which exists in the so-called
forbidden region where no bulk states occur. For the bulk
spectral density we take that of the center layer ($\alpha=10$) of the
20-layer film (see Fig.~\ref{figure3}).
The forbidden regions lie below and above the bulk bands 
but also includes ``white regions'' right in the middle of the bulk
bands. In this respect, in Fig.~\ref{figure3}, clearly the 
lowest bands of the spin-$\uparrow$ and of the spin-$\downarrow$ spectral 
densities of the surface layer ($\alpha=1$) around the $\Gamma$-point and
the M-point constitute surface state bands.
These surface states originate from the LSDA calculation. 
With the {\em s-f}\/ model calculation one can now investigate the
temperature dependence of the surface state. 

Fig.~\ref{figure4} shows the spectral density of the surface
layer of a 20-layer film at ${\bf k}=\bar{\Gamma}$ and ${\bf k}=\bar{\rm M}$
for different temperatures. 
For comparison, the respective spectral densities of the center layer of
the 20-layer films are plotted.
The plotted spectral densities clearly indicate surface state bands.
For $T=0$ ($\langle S_z \rangle/S=1$) the lower band edge of these
surface state bands lies at the $\Gamma$-point about $0.8\,{\rm eV}$ and
at the $\bar{\rm M}$-point about $0.45\,{\rm eV}$ below 
the lower conduction-band edge of the ``bulk'' bands of the central
layer (cf.~dashed line in Fig.~\ref{figure4}). These splittings between
the surface states and the lower edges of the bulk band are almost
independent on the temperature.  
With increasing temperature both the surface states and the lower band
edges of the bulk bands for the two spin directions converge Stoner-like. 

\begin{figure}[t!]
  \centerline{\includegraphics[width=\linewidth]{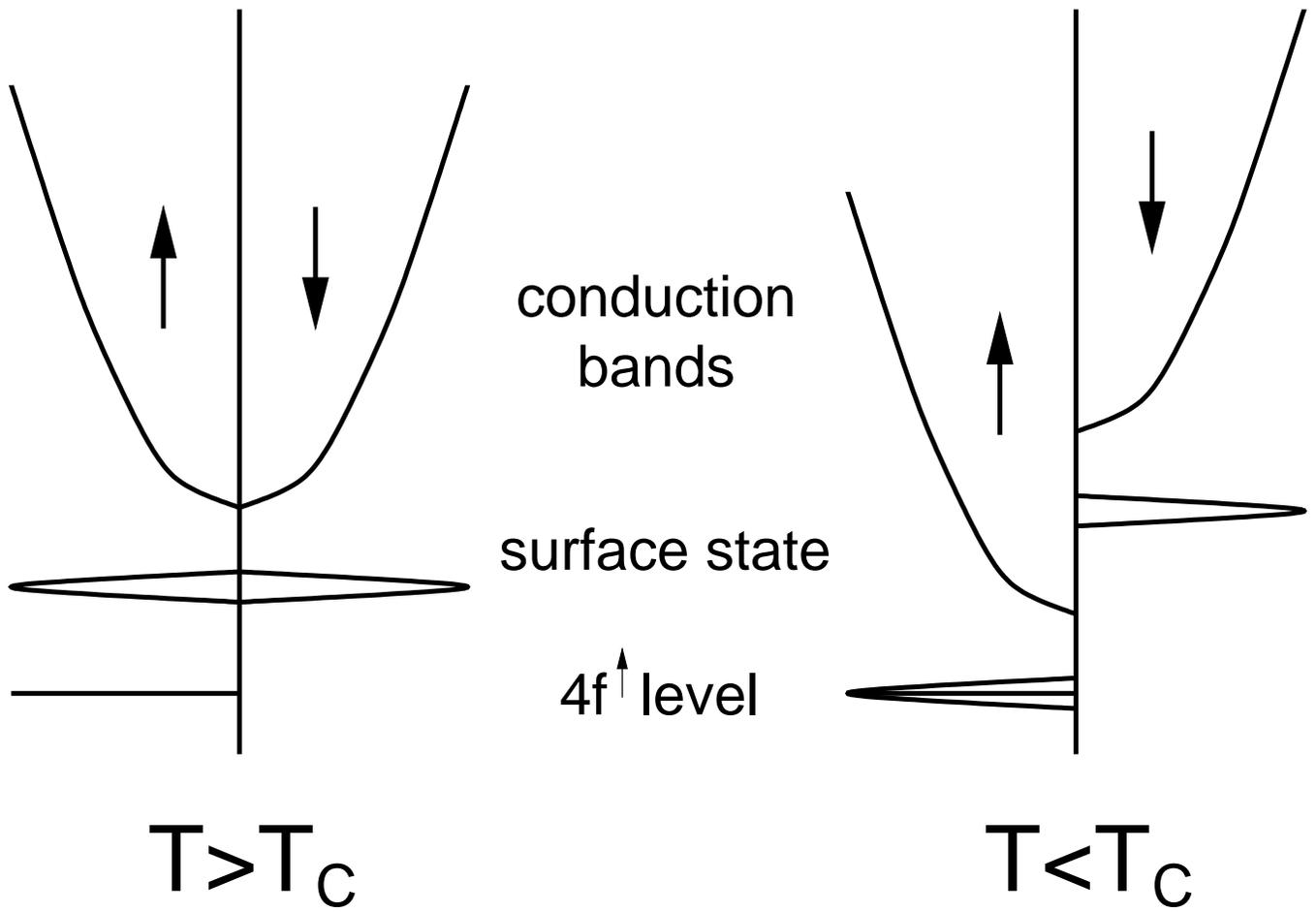}}  
  \caption{Schematics of the surface (half)metal-insulator-transition.
  For $T>T_{\rm C}$ the Fermi energy ($E_{\rm F}$) lies between the
  $4f^{\uparrow}$-levels and the surface state. For $T<T_C$ $E_{\rm F}$
  lies within the spin-$\uparrow$ surface state.}  
  \label{figure5}
\end{figure}

As a result, the surface state does not change the down shift of the
lower band edge upon cooling below $T_{\rm C}$ (red shift) of the 
system. However, the lower band edge of the LDOS at the surface is
lowered by about  
$0.8\,{\rm eV}$ compared to the bulk-like LDOS in the center of the film. 
Thus, for the 20-layer ${\rm EuO}(100)$ film, the band gap between
the occupied ${\it 4f}^{\uparrow}$ bands and the unoccupied 
${\it 5d}_{t_{2g}}$ bands will be reduced by the same $0.8\,{\rm eV}$.
When decreasing the temperature to $T=0$ the
red shift will reduce the band gap further: by $0.35\,{\rm eV}$ according
to our calculations and by $0.27\,{\rm eV}$ according to the experimental
results \cite{Wac91}. The overall reduction of the \linebreak[4]
{\em 4f}-${\it 5d}_{t_{2g}}$ gap will amount to $1.15\,{\rm eV}$ and 
$1.07\,{\rm eV}$, respectively. These values are exactly in the range of
the experimental {\em 4f}-${\it 5d}_{t_{2g}}$ gap of bulk 
${\rm EuO}$ at $300\,{\rm K}$ of $1.12\,{\rm eV}$ \cite{Wac91}.
These results indicate a possible surface insulator-metal transition in 
${\rm EuO}(100)$ films as a function of decreasing temperature, 
$T\rightarrow 0$ (cf.~Fig.~\ref{figure5}). Due to the exchange 
splitting of the conduction bands into the lower-energetic
spin-$\uparrow$ band and the higher-energetic spin-$\downarrow$ band the
respective phase for $T\rightarrow 0$ would be a half-metal.
As a result, the resistivity of the ${\rm EuO}(100)$ films should be
highly dependent on an applied magnetic field and a colossal
magnetoresistance (CMR) effect would be observed. 


We wish to acknowledge the financial support by the SFB 290 and by the
German National Merit Foundation. One of us (R.\,S.) is indebted to 
Prof.\ A.\ J.\ Freeman from Northwestern University for his hospitality
and many fruitful discussions.


\begin{thebibliography}{10}
\expandafter\ifx\csname bibnamefont\endcsname\relax
  \def\bibnamefont#1{#1}\fi
\expandafter\ifx\csname bibfnamefont\endcsname\relax
  \def\bibfnamefont#1{#1}\fi
\expandafter\ifx\csname url\endcsname\relax
  \def\url#1{\texttt{#1}}\fi
\expandafter\ifx\csname urlprefix\endcsname\relax\def\urlprefix{URL: }\fi
\providecommand{\bibinfo}[2]{#2}
\providecommand{\eprint}[2][]{\url{#2}}

\bibitem{DMIL97}
\bibinfo{author}{\bibfnamefont{P.~A.} \bibnamefont{Dowben}},
  \bibinfo{author}{\bibfnamefont{D.~N.} \bibnamefont{McIlroy}},
  \bibnamefont{and} \bibinfo{author}{\bibfnamefont{D.}~\bibnamefont{Li}}, in
  \emph{\bibinfo{booktitle}{Handbook of the Physics and Chemistry of Rare
  Earth}}, edited by \bibinfo{editor}{\bibfnamefont{K.~A.}
  \bibnamefont{{Gschneidner, Jr.}}} \bibnamefont{and}
  \bibinfo{editor}{\bibfnamefont{L.}~\bibnamefont{Eyring}}
  (\bibinfo{publisher}{Elsevier}, \bibinfo{address}{Amsterdam},
  \bibinfo{year}{1997}), vol.~\bibinfo{volume}{24}, chap.
  \bibinfo{chapter}{159}.

\bibitem{WAG+85}
\bibinfo{author}{\bibfnamefont{D.}~\bibnamefont{Weller}},
  \bibinfo{author}{\bibfnamefont{S.~F.} \bibnamefont{Alvarado}},
  \bibinfo{author}{\bibfnamefont{W.}~\bibnamefont{Gudat}},
  \bibinfo{author}{\bibfnamefont{K.}~\bibnamefont{Schr\"oder}},
  \bibnamefont{and} \bibinfo{author}{\bibfnamefont{M.}~\bibnamefont{Campagna}},
  \bibinfo{journal}{Phys. Rev. Lett.} \textbf{\bibinfo{volume}{54}},
  \bibinfo{pages}{1555} (\bibinfo{year}{1985}).

\bibitem{WF91}
\bibinfo{author}{\bibfnamefont{R.}~\bibnamefont{Wu}} \bibnamefont{and}
  \bibinfo{author}{\bibfnamefont{A.~J.} \bibnamefont{Freeman}},
  \bibinfo{journal}{J. Magn. Magn. Mater.} \textbf{\bibinfo{volume}{99}},
  \bibinfo{pages}{81} (\bibinfo{year}{1991}).

\bibitem{LHD+91}
\bibinfo{author}{\bibfnamefont{D.}~\bibnamefont{Li}},
  \bibinfo{author}{\bibfnamefont{C.~W.} \bibnamefont{Hutchings}},
  \bibinfo{author}{\bibfnamefont{P.~A.} \bibnamefont{Dowben}},
  \bibinfo{author}{\bibfnamefont{C.}~\bibnamefont{Hwang}},
  \bibinfo{author}{\bibfnamefont{R.-T.} \bibnamefont{Wu}},
  \bibinfo{author}{\bibfnamefont{M.}~\bibnamefont{Onellion}},
  \bibinfo{author}{\bibfnamefont{A.~B.} \bibnamefont{Andrews}},
  \bibnamefont{and} \bibinfo{author}{\bibfnamefont{J.~L.}
  \bibnamefont{Erskine}}, \bibinfo{journal}{J. Magn. Magn. Mater.}
  \textbf{\bibinfo{volume}{99}}, \bibinfo{pages}{85} (\bibinfo{year}{1991}).

\bibitem{DDN98}
\bibinfo{editor}{\bibfnamefont{M.}~\bibnamefont{Donath}},
  \bibinfo{editor}{\bibfnamefont{P.~A.} \bibnamefont{Dowben}},
  \bibnamefont{and} \bibinfo{editor}{\bibfnamefont{W.}~\bibnamefont{Nolting}},
  eds., \emph{\bibinfo{title}{Magnetism and Electronic Correlations in
  Local-Moment Systems: Rare-Earth Elements and Compounds}}
  (\bibinfo{publisher}{World Scientific}, \bibinfo{address}{Singapore},
  \bibinfo{year}{1998}).

\bibitem{SN99a}
\bibinfo{author}{\bibfnamefont{R.}~\bibnamefont{Schiller}} \bibnamefont{and}
  \bibinfo{author}{\bibfnamefont{W.}~\bibnamefont{Nolting}},
  \bibinfo{journal}{Solid State Commun.} \textbf{\bibinfo{volume}{110}},
  \bibinfo{pages}{121} (\bibinfo{year}{1999}).

\bibitem{Schiller00}
\bibinfo{author}{\bibfnamefont{R.}~\bibnamefont{Schiller}},
  \emph{\bibinfo{title}{Correlation Effects and Temperature Dependencies in
  Thin Ferromagnetic Films: Magnetism and Electronic Structure}},
  \bibinfo{type}{PhD thesis}, \bibinfo{school}{Humboldt-Universit\"at zu
  Berlin} (\bibinfo{year}{2000}),
  \urlprefix\url{http://dochost.rz.hu-berlin.de/dissertationen/abstract.php3?id=3000273}.

\bibitem{SN99b}
\bibinfo{author}{\bibfnamefont{R.}~\bibnamefont{Schiller}} \bibnamefont{and}
  \bibinfo{author}{\bibfnamefont{W.}~\bibnamefont{Nolting}},
  \bibinfo{journal}{Phys. Rev. B} \textbf{\bibinfo{volume}{60}},
  \bibinfo{pages}{462} (\bibinfo{year}{1999}).

\bibitem{quantmag2}
\bibinfo{author}{\bibfnamefont{W.}~\bibnamefont{Nolting}},
  \emph{\bibinfo{title}{Quantentheorie des Magnetismus, Teil 2: Modelle}}
  (\bibinfo{publisher}{Teubner}, \bibinfo{address}{Stuttgart},
  \bibinfo{year}{1986}).

\bibitem{FBS+93}
\bibinfo{author}{\bibfnamefont{M.}~\bibnamefont{Farle}},
  \bibinfo{author}{\bibfnamefont{K.}~\bibnamefont{Baberschke}},
  \bibinfo{author}{\bibfnamefont{U.}~\bibnamefont{Stetter}},
  \bibinfo{author}{\bibfnamefont{A.}~\bibnamefont{Aspelmeier}},
  \bibnamefont{and}
  \bibinfo{author}{\bibfnamefont{F.}~\bibnamefont{Gerhardter}},
  \bibinfo{journal}{Phys. Rev. B} \textbf{\bibinfo{volume}{47}},
  \bibinfo{pages}{11571} (\bibinfo{year}{1993}).

\bibitem{Wac91}
\bibinfo{author}{\bibfnamefont{P.}~\bibnamefont{Wachter}}, in
  \emph{\bibinfo{booktitle}{Alloys and Intermetallics}}, edited by
  \bibinfo{editor}{\bibfnamefont{K.~A.} \bibnamefont{{Gschneidner, Jr.}}}
  \bibnamefont{and} \bibinfo{editor}{\bibfnamefont{L.}~\bibnamefont{Eyring}}
  (\bibinfo{publisher}{Elsevier}, \bibinfo{address}{Amsterdam},
  \bibinfo{year}{1991}), vol.~\bibinfo{volume}{2} of
  \emph{\bibinfo{series}{Handbook of the Physics and Chemistry of Rare
  Earths}}, chap.~\bibinfo{chapter}{19}.

\bibitem{SMN97}
\bibinfo{author}{\bibfnamefont{R.}~\bibnamefont{Schiller}},
  \bibinfo{author}{\bibfnamefont{W.}~\bibnamefont{M\"uller}}, \bibnamefont{and}
  \bibinfo{author}{\bibfnamefont{W.}~\bibnamefont{Nolting}},
  \bibinfo{journal}{J. Magn. Magn. Mater.} \textbf{\bibinfo{volume}{169}},
  \bibinfo{pages}{39} (\bibinfo{year}{1997}).

\bibitem{And75}
\bibinfo{author}{\bibfnamefont{O.~K.} \bibnamefont{Andersen}},
  \bibinfo{journal}{Phys. Rev. B} \textbf{\bibinfo{volume}{12}},
  \bibinfo{pages}{3060} (\bibinfo{year}{1975 $r$}).

\bibitem{Fermi_energy}
\bibinfo{comment}{The Fermi energy level has been determined by substracting the experimental {\em 4f}-${\it 5d}_{t_{2g}}$ gap of bulk EuO at 300K of $1.\,12\,{\rm eV}$ \cite{Wac91} from the lower band edge of the center layer ($\alpha=10$) of the 20-layer film at $T=T_c$}.


\end{thebibliography}
\pagebreak

\end{document}